\documentclass[pra,twocolumn,showpacs,amsmath,amssymb]{revtex4}
\usepackage{graphicx}
\DeclareGraphicsExtensions{.png,.jpg,.eps}
\usepackage{dcolumn}
\usepackage{bm}
\usepackage[latin1]{inputenc}
\usepackage{xcolor}
\usepackage{float}
\usepackage{dsfont}
\usepackage{amsmath}

\begin{document}
\date{\today}

\title{Topological features of hydrogenated Graphene}
\author{Luis A. Gonzalez-Arraga$^1$, J. L. Lado$^2$ and Francisco Guinea$^{1,3}$}
\affiliation{$^1$IMDEA Nanociencia, Calle de Faraday, 9, Cantoblanco, 28049, Madrid, Spain}
\affiliation{$^2$International Iberian Nanotechnology Laboratory (INL), Av. Mestre Jose Veiga, 4715-330 Braga, Portugal}
\affiliation{$^3$School of Physics and Astronomy, University of Manchester, Oxford Road, Manchester M13 9PL, UK}

\begin{abstract}
Hydrogen adatoms are one of the most the promising proposals
 for the functionalization of graphene. Hydrogen induces narrow resonances near the Dirac energy, which lead to the formation
of magnetic moments. Furthermore, they also create local
lattice distortions which enhance the spin-orbit coupling. The combination of magnetism and spin-orbit coupling allows for a rich variety of phases, some of which have non trivial topological features. We analyze the interplay between magnetism and spin-orbit coupling in ordered arrays of hydrogen on graphene monolayers, and classify  the different phases that may arise. We extend our model to consider arrays of adsorbates in graphene-like crystals with stronger intrinsic spin-orbit couplings. 

\end{abstract}

\maketitle

\section{Introduction}

Adsorbates and vacancy effects in graphene have been a major field of research in recent years. An isolated vacancy
gives rise to localized resonance states near the Fermi level\cite{PGLPN06,UBGG10}. Hydrogen adatoms are expected to form a strong covalent bond with a carbon atom in the graphene lattice. The bond effectively removes one $\pi$ orbital in the graphene band, leading to a sharp resonance near the Fermi energy, in a similar way to the case of the vacancy\cite{BKL08}. The electron-electron repulsion prevents doubly occupied states of this resonance, and lead to the formation of a magnetic moment.  In addition, the carbon atom coupled to the hydrogen atom is displaced from the graphene plane, inducing a local sp$^3$ hybridization, which increases the spin-orbit coupling\cite{NG09,GKF13}.

The intrinsic spin-orbit coupling in a perfect graphene layer creates a gap, and turns graphene into a topological insulator\cite{KM05}. The presence of magnetic moments induces a exchange coupling with the spins of itinerant electrons, and breaks time reversal symmetry. The combination of a uniform exchange coupling and the extrinsic Rashba coupling in graphene leads to a quantum anomalous Hall phase\cite{Qetal10}. This phase is an example of systems which do not show time reversal symmetry and have topologically protected edge states without Landau levels\cite{H88}.

The effect of a uniform magnetic field and the spin-orbit coupling has been extensively studied in silicene\cite{E12,GBT13,Hetal14}. The possibility of inducing non trivial topological features in the electronic structure of graphene by the addition of heavy atoms has also been considered\cite{Wetal11,HAWF12,Cetal14}, and non trivial features associated to the spin-orbit coupling have been found in lead intercalated CVD graphene\cite{Cetal15}.

On one hand, the formation of magnetic moments near vacancies and hydrogen atoms in graphene has been extensively investigated\cite{KUPGN12},  both theoretically (see, for instance \cite{OL07,PFB08,HFPOG11,PS14,Getal14}), and experimentally\cite{Zetal09,Setal10,WOBZ10,Netal12,Cetal12}.
On the other hand, it is known that
adsorbates on graphene form ordered arrays in a variety of situations\cite{Betal10,MR14}. Adatoms on graphene interact among themselves, and can form a variety of ordered patterns\cite{SAL09,CSAF10,ASL10}. More importantly, as shown recently, hydrogen adsorbates can be manipulated with an STM tip\cite{B15}.

We analyze here the combined effects of the exchange coupling and modified spin-orbit coupling due to hydrogen adsorbates which form a regular array. As mentioned before, the formation of ordered arrays can be induced by the interactions between adatoms, or they can be fabricated on purpose. The results are expected to give hints on the local changes in the electronic structure near an isolated impurity. In section II we present the model, and discuss the main features of arrays of hydrogen adatoms. In section III we study the topological properties of the hydrogenated graphene superlattice considering only the local adatom-induced SO couplings. In section IV we proceed to study related situations, where the spin-orbit coupling throughout the entire lattice cannot be neglected. Finally, we discuss the most relevant results, and the open questions raised by our work.

\section{The model.}
\begin{figure}[h]\label{fig1}
\begin{center}
\includegraphics[width=8cm]{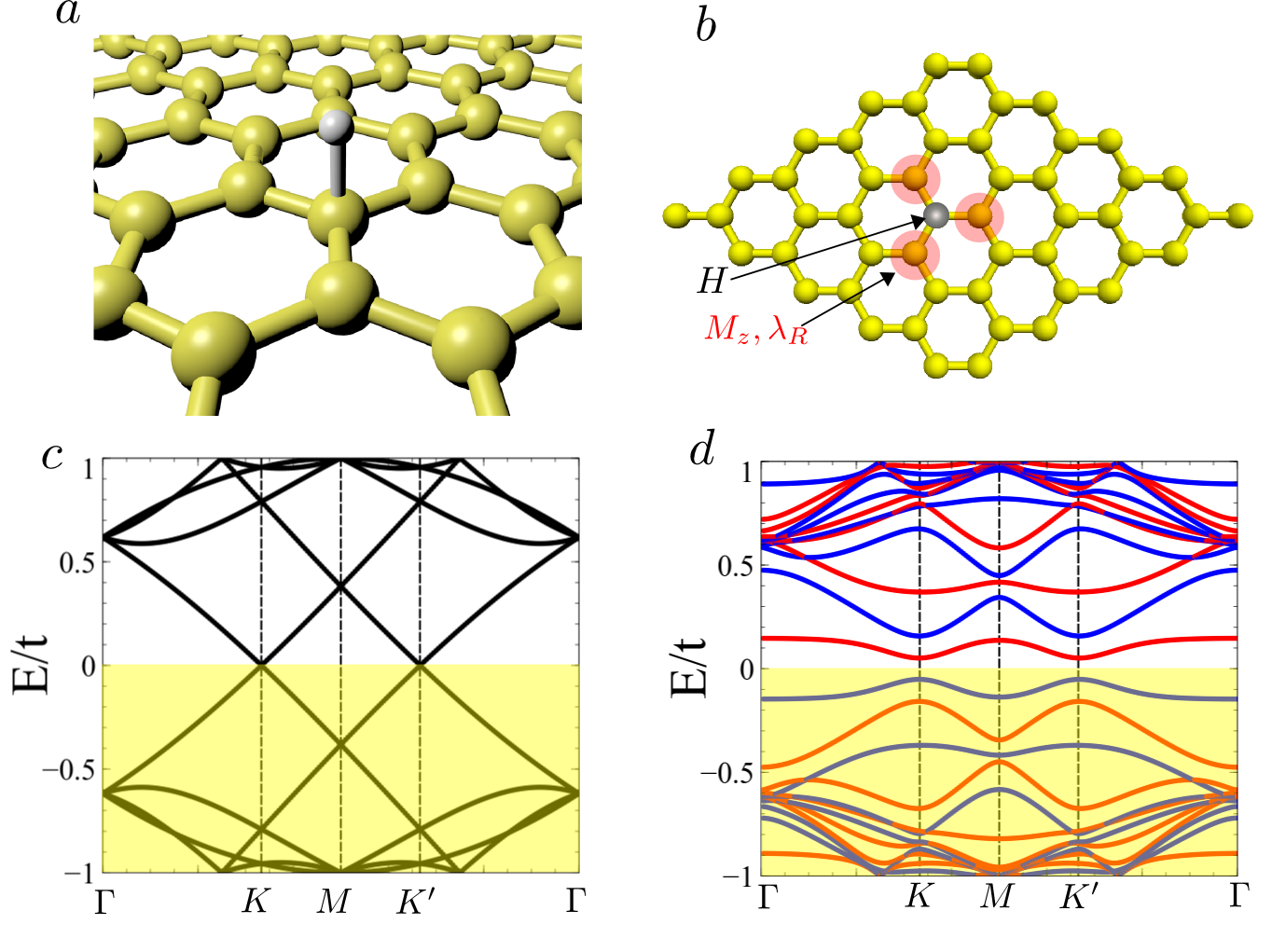}
\caption{ Single hydrogen adatom in  graphene sheet (a), and unit cell
of the simulation (b). As shown in (b), Rashba and exchange fields
are located only around the adatom. Band structure of a pristine 5x5
graphene unit cell (c), and with a vacancy and local exchange field (d).
}
\end{center}
\end{figure}

We will study the topological properties of a periodic array composed of a  hydrogen adatom on top of a carbon atom in the $5\times 5$ supercell
of graphene as shown in the first two panels of  \ref{fig1} . We describe the electronic structure with a tight binding model with one $p_z$ orbital per carbon atom and a nearest neighbor hopping parameter, $t$. The effect of the covalent bonding between hydrogen and carbon is approximated by a large shift of the energy of the $p_z$ orbital nearest to the hydrogen atom, $\epsilon_0$. For $\epsilon_0 \gtrsim | t |$ a sharp resonance appears near the vacancy\cite{PGLPN06}, which decays slowly as function of the distance. The resonance is mostly localized in the sublattice which does not include the perturbed $p_z$ orbital.

We assume that the electron-electron interaction induces a exchange splitting, $\Delta_Z$ between the spin up and spin down levels at the sites where the resonance is located. This is the effect which arises if the interaction is local and it is treated within a mean field approximation. The orientation of the exchange field with respect to the plane of the graphene layer is determined by the spin-orbit coupling. To a first approximation, a Rashba-like coupling does not modify the spin, as a Rashba coupling implies hopping between different sublattices and the spin is localized in one sublattice (note that the itinerant electrons are affected). The full Hamiltonian for the adatom in the supercell is:
\begin{equation}
H=H_{KIN}+H_{Z}+H_{SO}
\end{equation}
where $H_{KIN}$ represents the kinetic terms of the Hamiltonian, $H_{SO}$ represents the spin-orbit coupling terms, and $H_{Z}$ is the exchange splitting due to the magnetism induced by the onsite potential of the adatom, which takes the form of a usual site dependent Zeeman term. In this section we will explain the kinetic and exchange contributions to the Hamiltonian. In the next two sections we will discuss two cases  separately: local SO coupling induced by the adatom, and uniform SO coupling induced by a substrate. 

The kinetic term can be written:

\begin{equation}
H_{KIN}=-t\sum_{\langle i,j \rangle,s}c^{\dagger}_{is}c_{js}+\epsilon_{o}\sum_{s}c^{\dagger}_{C_H, s}c_{C_H, s}
\end{equation}

where  $t$ is the hopping between
nearest neighbors $i$ and $j$ in the honeycomb lattice, $\epsilon_{0}$ is a large onsite potential in the hydrogenated carbon site (in our calculation we consider it to be $100$ times the hopping integral), and $c^{\dagger}_{C_H, s}$ ($c_{C_H, s}$) creates (annihilates) an  electron
in the hydrogenated carbon site ($C_H$). The onsite potential due to the adatom gives rise to a gap between the conduction and valence bands, as well as a vacancy state lying at
the Fermi energy.  The size of this gap is proportional to the impurity concentration, so that in the dilute limit of an isolated vacancy in graphene, there is a zero energy vacancy state, but no gap.

\begin{figure}[h]\label{fig2}
\begin{center}
\includegraphics[width=8cm]{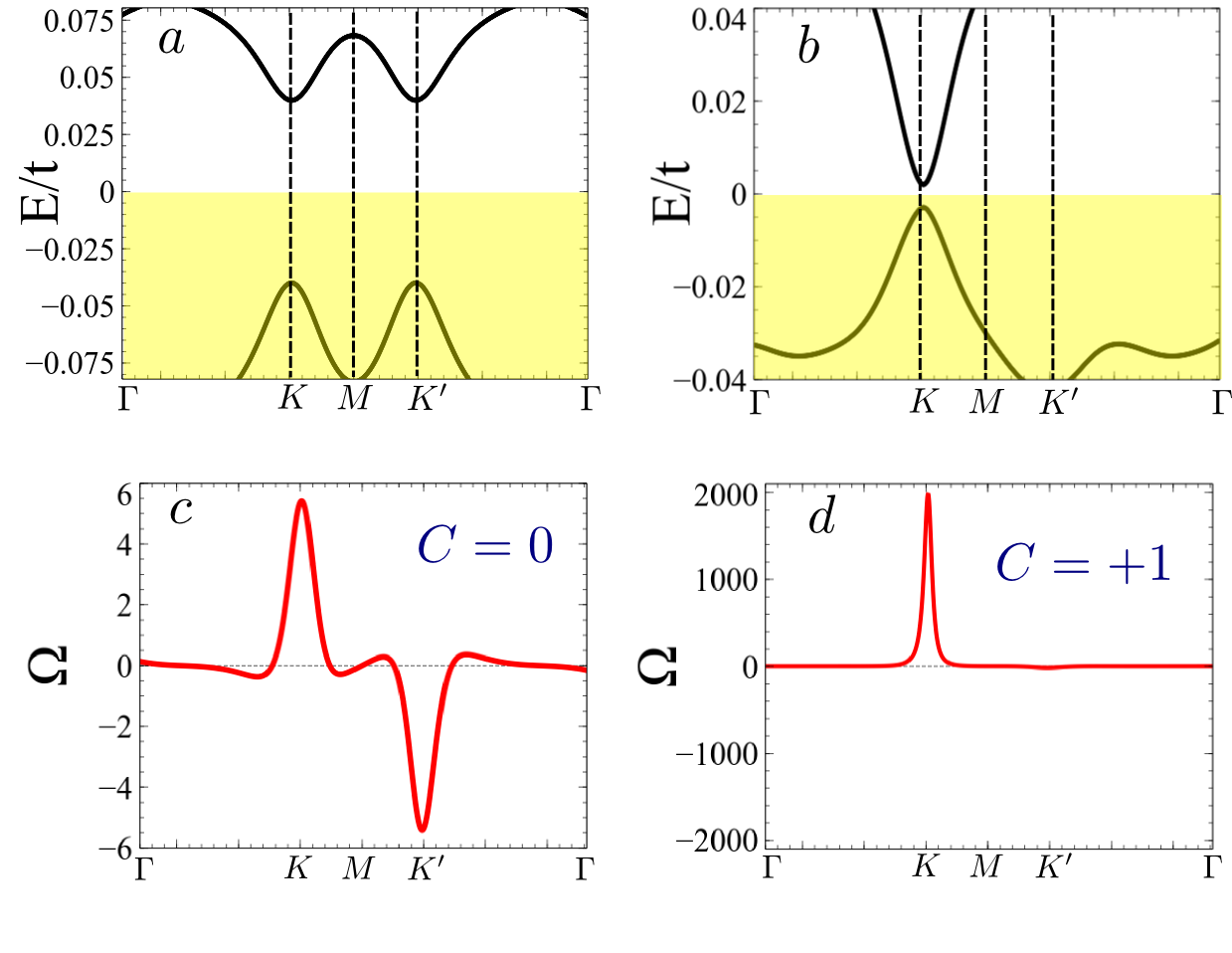}
\caption{Band structures (a,b) and Berry curvatures (c,d) of a graphene supercell
with one hydrogen atom, with local Rashba field and local off-plane exchange. Panels
(a,c) corresponds to realistic values of $\lambda_R$, whereas panels (b,d)
corresponds to a large $\lambda_R$.}
\end{center}
\end{figure}

The vacancy state also produces a non-uniform internal magnetization in the sub lattice opposite to the adatom, which decreases with distance from the impurity site. The local magnetization splits the impurity bands, creating a gap at the Fermi level, which can be seen in panel (d) of \ref{fig1}.  We model this magnetization effect by the exchange splitting term in the Hamiltonian, which has the form:
\begin{equation}\label{zeeman}
H_{Z}=\sum_{i_B }\Delta_{i_B} c^{\dagger}_{i_B, s} s_{z} c_{i_B, s}
\end{equation}
where the $i_{B}$ label means that the summation is carried out over all sites of the B sub lattice opposite to $C_{H}$. Effectively, this term contributes a spin-dependent onsite potential in the B sites. The splitting magnitude $\Delta_{i_B}$ is proportional to the magnitude of the vacancy state at site $i_{B}$, that is:
\begin{equation}
\Delta_{i_B}=\Delta \vert \Psi_v (i_{B})\vert ^2
\end{equation}
After solving for the band structure of the spin-unpolarized kinetic part of the Hamiltonian,  we introduce the sub lattice magnetization via eq. (\ref{zeeman}).

\section{HYDROGENATED GRAPHENE: LOCAL SPIN-ORBIT COUPLINGS}

In this section we discuss the model with local, adatom-induced SO couplings. The Hamiltonian for the spin orbit coupling induced by the adatom is \cite{GKF13}:
\begin{eqnarray}
H_{SO}&=&\\ \nonumber
&&\frac{i}{3} \sum_{\left\langle \left\langle C_H , j \right\rangle \right\rangle} c^{\dagger}_{C_H, s} c_{j s^{\prime}} \left[  \frac{\Lambda_I }{\sqrt{3}} \nu_{C_H , j} s_z     \right]_{s s^\prime} +H.c. \\ \nonumber
&+&     \frac{2i}{3} \sum_{ \left\langle \left\langle C_H, j \right\rangle \right\rangle} c^{\dagger}_{C_H, s} c_{nn,j, s^{\prime}}  \left[  \lambda_{R} (\vec{s} \times \hat{d_{C_H,j}} )_z     \right ]_{s s^\prime} +H.c.        \\ \nonumber
&+& \frac{2i}{3} \sum_{\left\langle \left\langle i,j \right\rangle \right\rangle}c^{\dagger}_{nn,i, s} c_{nn,j, s^{\prime}}  \left[  \Lambda_{PIA} (\vec{s} \times \hat{D_{i,j}} )_z     \right ]_{s s^\prime}
\end{eqnarray}
Where $\vec{d}_{C_H,j}$ is the unit vector connecting $C_H$ to its nearest neighbors, and  $\vec{D}_{i,j}$  is the unit vector
connecting the second neighbors of $C_H$.  The operator $c^{\dagger}_{nn,j,s}$ ( $c_{nn,j,s}$) creates (annihilates) electrons of
spin $s$ in the $j$th nearest neighbor of $C_{H}$, whereas $c^{\dagger}_{C_H,s}$ creates an electron at $C_H$.The first term
describes the adatom modified intrinsic SO coupling $\Lambda_{I}$, which induces spin-preserving hoppings
between $C_H$ and its second neighbors.  The second term ($\lambda_{R}$) describes the Bychkov-Rashba hoppings caused by the
local breaking of space inversion symmetry around the adatom site. It induces spin-flipping hoppings between $C_{H}$ and
its nearest neighbors. The third term ($\Lambda_{PIA}$) induces spin-flipping hoppings between the nearest
neighbors of $C_{H}$. We have neglected in this section the intrinsic SO coupling of pristine graphene, which induces hoppings
between second neighbors not containing $C_{H}$.

The presence of Rashba interaction and an exchange field in the sites where the impurity bands are located,
strongly suggests the possibility of reaching a Quantum Anomalous Hall (QAH) state, via a topological phase transition 
occurring in the corresponding gap. We inspect the topological properties of the valence bands of the bulk
system, by calculating the Chern number
\begin{equation}
\mathcal{C}=\frac{1}{2\pi}\sum_{n} \int_{BZ} \Omega_{n}(k_x ,k_y ) d^2 k
\end{equation}
where the integral is carried out over the first Brillouin zone, and the summation is carried out over all valence bands. The Chern number gives the number of chiral edge states at each edge of a nano ribbon, and is related to the quantized charge Hall conductance via $\sigma_{yx}=\mathcal{C} e^2 / \hbar$ . Using the ab-initio values reported by \cite{GKF13}  for the spin-orbit coupling strengths and onsite potential, we calculate
the Berry curvature of the graphene supercell with an adatom. The SO coupling strengths
are $\lambda_{R}=1.14\times 10^{-4}$ $t$, $\Lambda_{PIA}=-2.66\times 10^{-4}$ $t$ and $\Lambda_{I}=-7.26\times 10^{-4}$ $t$. The
magnitude of the exchange splitting at the nearest neighbors of the adatom site is set at $0.1$ $t$ throughout
this calculation. A close-up of the band structure around the gap, and the Berry curvature profiles for these values are shown in panels (a) and (c) of  Fig. (2), integration over the Brillouin zone yields $\mathcal{C}=0$. The Berry curvature has different signs in
opposite valleys and the Chern number around each valley is not an integer. This situation is similar to the Valley Hall
effect in graphene in the presence of a sub lattice staggered potential, where valley currents are formed in the direction
transverse to an applied in-plane electric field.

Moving away from graphene, we will
explore this model in the regime of stronger SOC, which
will be representative of heavier graphene like honeycomb lattices.
We will investigate the conditions for achieving a QAH state in this setting,  we do this by checking the
low-energy band structure and Berry curvature while increasing the Rashba parameter. A closing and reopening of the gap occurs at
the topological phase transition from the Valley-Hall regime to the QAH state. 
We found this to occur for values of the 
Rashba parameter that are $10^4$ times larger than the ab-initio values of \cite{GKF13} ($\lambda_{R}=3.25$ $t$). Panel (b) in Fig. (2) shows  the band
structure for this situation of very large Rashba coupling, whereas panel (d) shows the profile of the Berry curvature in the same situation. The magnetization breaks both  time reversal symmetry and  sub lattice symmetry, since the adatoms are located only in one sublattice. We find the Berry curvature to be almost entirely concentrated
around the $K$ valley. We find $\mathcal{C}=1$. We have investigated the band gaps at half-filling
for lower adatom concentrations, and we have found the QAH phase to exist within a region of parameter
space beyond physically realistic values\cite{notePb}. Therefore,
although theoretically the simultaneous existence of exchange and Rashba
couplings in graphene might be able to create a quantum anomalous Hall state,
within the realistic values such phase is not expected to be observed.
Nevertheless, it is worth to note that the present model
can be also applied to model further 2D honeycomb systems
with stronger
SOC, in which the topological state could be realized. In the
following section, we will show how in these systems,
where apart form the Rashba coupling there is a strong
intrinsic SOC, the topological state can be easily realized.

\section{Honeycomb lattice with adsorbates and uniform SOC}

\begin{figure}[h]
\begin{center}
\includegraphics[width=8cm]{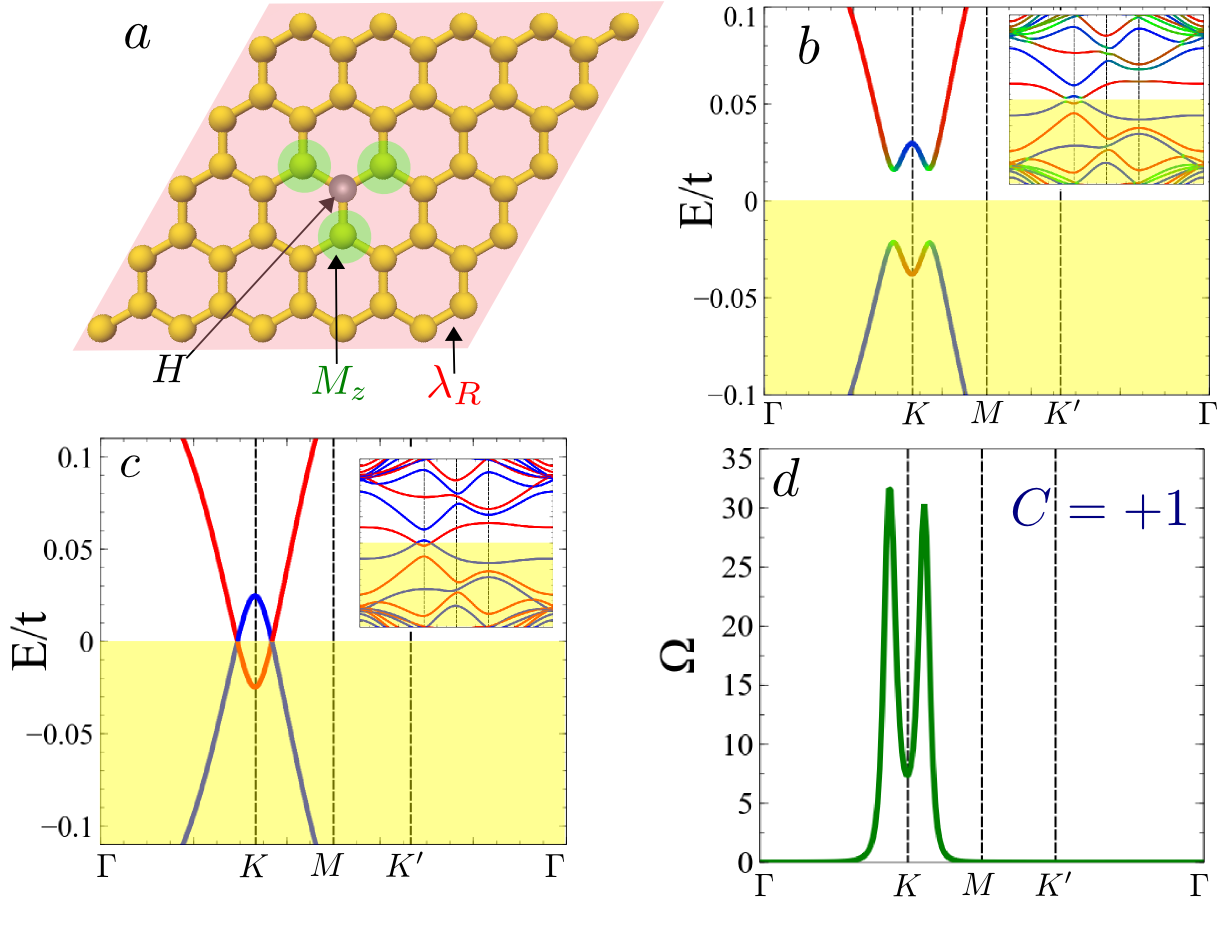}
\caption{ Scheme (a) of unit cell with vacancy induced local magnetism,
and uniform Rashba induced by off-plane electric field. Band structure
with vacancy, local magnetism,  uniform intrinsic SOC with (b) and
without Rashba (c). Berry curvature for (b), showing a Chern number
$C=+1$. In comparison with Fig.2, the anomalous Hall state (b) appears
at arbitrary small Rashba field, provided intrinsic SOC closes
the band gap (c).
}
\end{center}
\end{figure}

We have seen that the hydrogen-induced local Rashba is not strong enough 
to induce a topological phase transition to a 
QAH insulator, now we investigate the possibility of achieving 
such a state in a different setting: a supercell of a graphene like honeycomb crystal in which the intrinsic SO coupling cannot be neglected, in the presence of a substrate that induces a uniform Rashba effect, see panel (a) in Fig. (3). This model can be applied to a large family of systems
such as silicene, germanene, stanene, hydrogenated bismuth, metal
organic frameworks  \cite{RE13,MHSSKM2006,YYQZF2007,KGF2010,LFY2011,WLL2013} .In this model, the role of the adsorbate impurity is to generate the flat impurity band and to induce a non-uniform exchange field.  The spin-orbit coupling Hamiltonian in this model has the form:
\begin{eqnarray}
H_{SO}&=&\\ \nonumber
&+&     \frac{2i}{3} \sum_{\langle i, j \rangle} c^{\dagger}_{i s} c_{j s^{\prime}}  \left[  \lambda_{R} (\vec{s} \times \hat{d_{i,j}} )_z     \right ]_{s s^\prime} \\ \nonumber
&+&    \frac{i}{3} \sum_{\langle \langle i, j \rangle \rangle} c^{\dagger}_{i s} c_{j s^{\prime}} \left[  \frac{\Lambda_I }{\sqrt{3}} \nu_{i , j} s_z     \right]_{s s^\prime} +H.c.
     \\ \nonumber
\end{eqnarray}
where the SO hoppings are now uniformly distributed around the supercell. Apart from this, the supercell Hamiltonian remains
unchanged. Since the magnitude of the gap between the impurity bands decreases with adatom concentration, we  can consider a regime in which the Kane-Mele intrinsic SO coupling is strong enough to close the
exchange-splitted gap, as shown in panel (c) of Fig. (3). This situation can be observed with $\Lambda_I=0.02$ $t$, $\lambda_{R}=0.0$ $t$, and a exchange
splitting of $0.25$ $t$ in the nearest-neighbors of the adatom.  There is no gap around the K valley in this regime. Once the
Rashba SO coupling is induced, a gap with topologically non-trivial properties appears as shown in panels (b) and (d)
of Fig. (3), for the band structure and Berry curvature respectively,  for a Rashba of  $\lambda_{R}=0.02$ $t$.  Integration yields a Chern number $\mathcal{C}=1$, signaling a QAH state with 1 edge state.

A qualitative description of the properties of the system can be seen from the phase diagram in panel (a) of Fig. (4).  For small values of the intrinsic and Rashba SO couplings the system is a trivial insulator. By increasing the values of either the Kane-Mele or Rashba the system turns into a metal, whereas the combination of higher intrinsic and Rashba couplings changes the system into a QAH insulator with one conducting edge state. The regime where the system is a trivial insulator, is characterized by  Berry curvatures of opposite signs in opposite valleys.

So far, we have considered the topological properties of the superlattice of adsorbs at half-filling. The adsorbates break sub lattice inversion symmetry, and in the trivial insulating case, they generate a Berry curvature distribution similar to that of graphene in the presence of a sublattice staggered potential, characterized by peaks of opposite signs at the two valleys. When the Fermi level is tuned to be at the bottom of the conduction band, which can be done by applying a gate voltage, the Berry curvature distribution continues to be asymmetric in opposite valleys, as shown in panels  (c) and (d) of Fig. (4)., giving rise to a non-quantized Valley-Hall effect, i.e. skew-scattering for electrons in opposite valleys (see panel b), and thus a net valley current that travels in the bulk. The sign of the anomalous velocity can be switched by changing the sub lattice in which the adsorbates are placed.

This later effect, independent of the Rashba coupling,
is the one to be observed in graphene, provided all the hydrogen
impurities are located in the same sublattice. Nevertheless,
in a realistic random
hydrogenation, provided clustering can be avoided \cite{Gargiulo},
in a simplistic picture it is expected
that the hydrogen atom will lie randomly in both sublattices,
therefore giving rise to vanishing valley Hall effect. However,
if all the hydrogen atoms are located in the same sublattice,
the net valley Hall current will be different from zero and
strong non-local signal might be observed, very much like in
gated bilayer graphene \cite{SYBWTT15}.

\begin{figure}[h]
\begin{center}
\includegraphics[width=8cm]{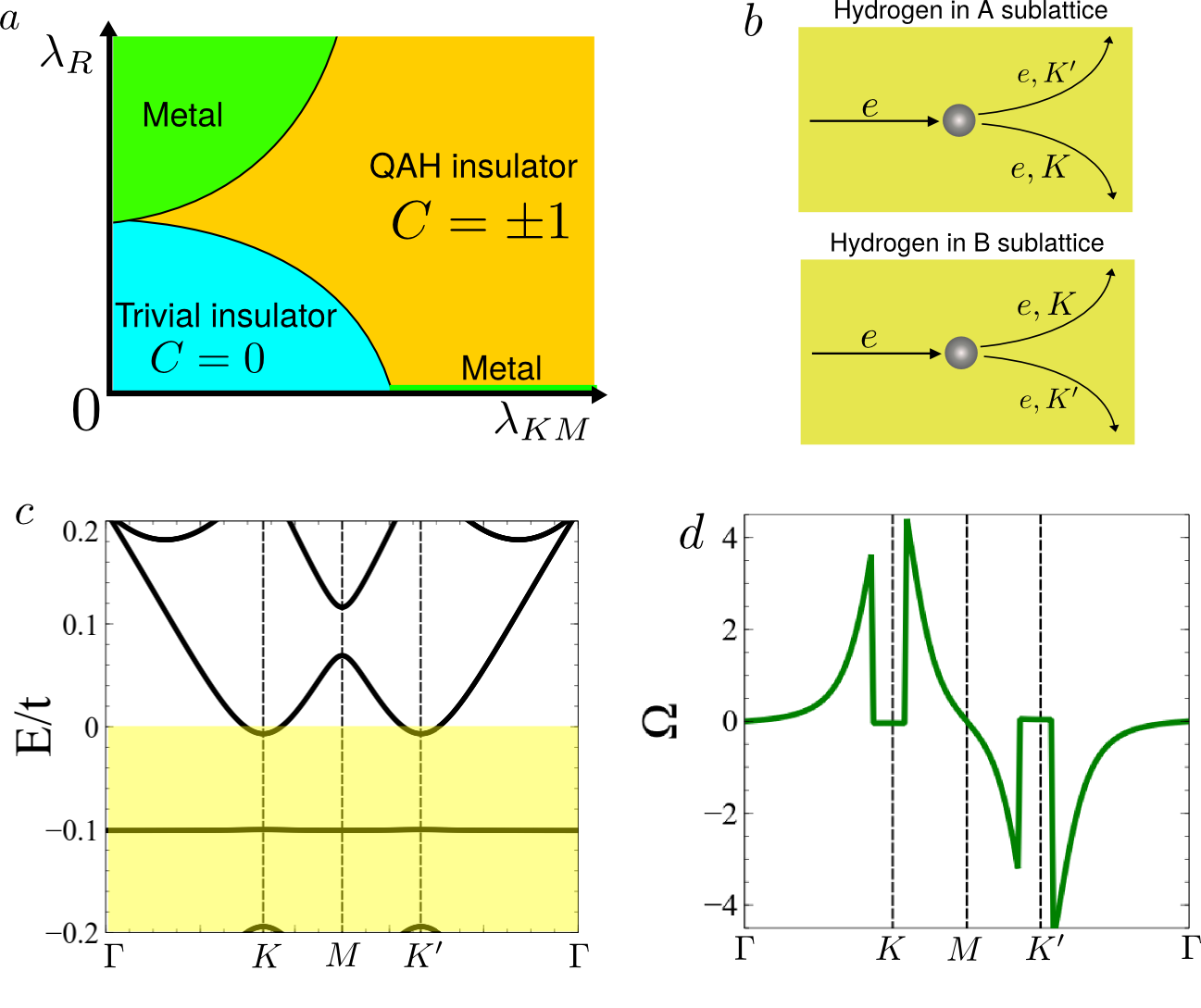}
\caption{(a) Schematic phase diagram of the adatom in the supercell with intrinsic and Rashba SO couplings as parameters, the exact values of the SO couplings vary as a function of impurity concentration. (b) The Berry curvature is concentrated in one valley or the other depending on which sub lattice the impurities are located. Valley currents of opposite signs may be thus be engineered at will. (c)  and (d) show respectively, the band structure and Berry curvature
for slightly doped graphene. If the Fermi level now lies at the bottom of the conduction band the Berry curvature still has opposite signs at opposite valleys.}
\end{center}
\end{figure}

\section{SUMMARY AND DISCUSSION}

We have aimed at an understanding of the topological properties of hydrogenated graphene in the dilute limit by a Berry curvature analysis of  the bulk gap at half filling. This gap is opened by the splitting of the impurity bands due to the exchange field. It is influenced by adatom concentration inasmuch as the vacancy state and the induced magnetic moment depend on the distance between adatoms \cite{soriano}. We have found that within the range of realistic values for the SO couplings and onsite potential reported in \cite{GKF13} the sub lattice asymmetry induced by the adatom predominates over the exchange field and local SO couplings. The predominant phase is similar to the Valley Hall effect in gapped graphene, where electrons in opposite valleys acquire anomalous velocities transverse to an in-plane electric field in opposite directions, generating thus a valley current that travels in the bulk. We have also found that this system can theoretically undergo  a topological phase transition to a valley-polarized Quantum Anomalous Hall state, but this QAH state lives in a region of parameter space with Rashba coupling values that are $10^4$ times larger than the ab-initio values reported by \cite{GKF13}.

We also extended our model in order to study the topological properties of arrays of adsorbates in honeycomb crystals in which the uniform intrinsic and Rashba SO couplings cannot be neglected, such as stanene, silecene or germanene. Starting from a situation in which the intrinsic SOC is strong enough to mix the exchange-splitted impurity bands, a situation that can be realized for low concentrations of impurities, we have found that the Rashba parameter is capable of opening a gap with non-trivial topological properties, effectively turning the system into a QAH insulator.  Finally we also considered the situation in which the Fermi level lies inside the bottom of the conduction band, we have found a Berry curvature distribution of peaks of opposite signs in the two valleys, marking the the presence of bulk valley currents in a direction transverse to an applied electric field.

\section{Acknowledgments}
We thank H. Ochoa, D. Bandurin and J. Fernandez-Rossier for helpful conversations. We acknowledge financial support by Marie-Curie-ITN 607904-SPINOGRAPH. LAGA and FG acknowledge support from the
Spanish Ministry of Economy (MINECO) through Grant
No. FIS2011-23713, the European Research Council Advanced
Grant (contract 290846), and the European Commission
under the Graphene Flagship, contract CNECTICT-
604391.


\begin{thebibliography}{99}
\bibitem{notePb}
Note, however, that spin-orbit coupling terms of order $\sim 1$ eV, although not of the Rashba type, have been reported for Pb intercalated CVD graphene\cite{Cetal15}.
\end{thebibliography}
\end{document}